\documentclass{aa}

\usepackage[nolist,nohyperlinks]{acronym}

\usepackage{siunitx}
\usepackage{microtype}

\usepackage{bm}

\usepackage{enumitem}

\usepackage{widetext}

\usepackage{lipsum}

\usepackage{silence}
\usepackage[colorlinks, allcolors=blue, breaklinks=true]{hyperref}
\usepackage{cleveref}

\usepackage{xcolor}

\usepackage{amsmath}

\usepackage{changepage}

\usepackage{tabularx}

\usepackage{tikz}
\usetikzlibrary{shapes,arrows,fit,positioning,calc}

\let\endlongtable\relax

\usepackage[rightcaption, ragged]{sidecap}
\sidecaptionvpos{figure}{t}

\usepackage{tabularx}

\usepackage{txfonts}

\usepackage[encapsulated]{CJK}
\usepackage{ucs}

\newcommand{\orcid}[1]{\protect\href{https://orcid.org/#1}{\protect\includegraphics[width=8pt]{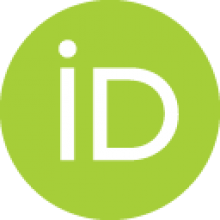}}}

\usepackage{xspace}

\def\M87{M87$^*$\xspace}
\def\m87{M87$^*$\xspace}
\def\sgra{Sgr~A$^*$\xspace}
\def\3C279{3C\,279\xspace}
\def\3c279{3C\,279\xspace}
\def\NRAO530{NRAO\,530\xspace}
\def\nrao530{NRAO\,530\xspace}
\def\J1924{J1924-2914\xspace}
\def\j1924{j1924-2914\xspace}
\def\z{\mbox{\textsc{Zingularity}}\xspace}
\def\symba{\mbox{\textsc{Symba}}\xspace}
\def\rpicard{\mbox{\textsc{Rpicard}}\xspace}
\def\meqsilhouette{\mbox{\textsc{MeqSilhouette}}\xspace}

\begin{document}

\begin{acronym}
\acro{sn}[S/N]{signal-to-noise ratio}
\acro{vlbi}[VLBI]{very long baseline interferometry}
\acroplural{vlbi}[VLBI]{Very-long-baseline interferometry}
\acro{grmhd}[GRMHD]{general relativistic magnetohydrodynamics}
\acro{grrt}[GRRT]{general relativistic ray-tracing}
\acro{mhd}[MHD]{magnetohydrodynamics}
\acro{as}[as]{arcseconds}
\acro{agn}[AGN]{Active Galactic Nuclei}
\acro{llagn}[LLAGN]{low-luminosity AGN}
\acro{riaf}[RIAF]{radiatively inefficient accretion flow}
\acro{cena}[Cen~A]{Centaurus~A}
\acroplural{cena}[Cen~A, NGC~5128]{Centaurus~A}
\acro{sgra}[Sgr\,A*]{Sagittarius~A*}
\acro{m87}[\m87]{Messier 87*}
\acro{agn}[AGN]{active galactic nuclei}
\acro{eht}[EHT]{Event Horizon Telescope}
\acro{bhc}[BHC]{\href{https://blackholecam.org}{BlackHoleCam}}
\acro{tanami}[TANAMI]{Tracking Active Galactic Nuclei with Austral Milliarcsecond Interferometry}
\acro{smbh}[SMBH]{supermassive black hole}
\acroplural{smbh}[SMBHs]{supermassive black holes}
\acro{aa}[ALMA]{Atacama Large Millimeter/submillimeter Array}
\acro{ap}[APEX]{Atacama Pathfinder Experiment}
\acro{pv}[PV]{IRAM~30\,m Telescope}
\acro{jc}[JCMT]{James Clerk Maxwell Telescope}
\acro{lm}[LMT]{Large Millimeter Telescope Alfonso Serrano}
\acro{sp}[SPT]{South Pole Telescope}
\acro{sm}[SMA]{Submillimeter Array}
\acro{az}[SMT]{Submillimeter Telescope}
\acro{c}[c]{speed of light}
\acro{pc}[pc]{parsec}
\acro{gr}[GR]{general relativity}
\acro{aips}[\textsc{aips}]{\href{http://www.aips.nrao.edu}{Astronomical Image Processing System}}
\acro{casa}[\textsc{casa}]{\href{https://casa.nrao.edu}{Common Astronomy Software Applications}}
\acro{symba}[\textsc{Symba}]{\href{https://bitbucket.org/M_Janssen/symba}{SYnthetic Measurement creator for long Baseline Arrays}}
\acro{rpicard}[\textsc{Rpicard}]{\href{https://bitbucket.org/M_Janssen/picard}{Radboud PIpeline for the Calibration of high Angular Resolution Data}}
\acro{hops}[\textsc{hops}]{\href{https://www.haystack.mit.edu/tech/vlbi/hops.html}{Haystack Observatory Postprocessing System}}
\acro{hbt}[HBT]{Hanbury Brown and Twiss}
\acro{tov}[TOV]{Tolman-Oppenheimer-Volkoff}
\acro{mri}[MRI]{magnetorotational instability}
\acro{adaf}[ADAF]{advection-dominated accretion flow}
\acro{adios}[ADIOS]{adiabatic inflow-outflow solution}
\acro{cdaf}[CDAF]{convection-dominated accretion flow}
\acro{bz}[BZ]{Blandford-Znajek}
\acro{bp}[BP]{Blandford-Payne}
\acro{em}[EM]{electromagnetic}
\acro{blr}[BLR]{broad-line region}
\acro{nlr}[NLR]{narrow-line region}
\acro{ism}[ISM]{interstellar medium}
\acro{edf}[eDF]{electron distribution function}
\acro{pic}[PIC]{particle-in-cell}
\acro{sane}[SANE]{standard and normal evolution}
\acro{mad}[MAD]{magnetically arrested disk}
\acro{jive}[JIVE]{\href{http://www.jive.eu}{Joint Institute for VLBI ERIC}}
\acro{nrao}[NRAO]{\href{https://www.nrao.edu}{National Radio Astronomy Observatory}}

\acro{muas}[$\mu$as]{microarcseconds}
\acroplural{agn}[AGN]{Active galactic nuclei}
\acro{jy}[Jy]{Jansky}
\acro{pa}[PA]{position angle}
\acro{srmhd}[SRMHD]{special relativistic magnetohydrodynamics}
\end{acronym}

\title{Deep learning inference with the Event Horizon Telescope}
\subtitle{III. \z{} results from the 2017 observations and \\ predictions for future array expansions}

   \author{M. Janssen\orcid{0000-0001-8685-6544}
          \inst{1,2}
        \and C.-k. Chan\orcid{0000-0001-6337-6126}\inst{3,4,5}
        \and J. Davelaar\orcid{0000-0002-2685-2434}\inst{6,7}
        \and M.~Wielgus\orcid{0000-0002-8635-4242}\inst{8}
          }
    \institute{Department of Astrophysics, Institute for Mathematics, Astrophysics and Particle Physics (IMAPP), Radboud University, P.O. Box 9010, 6500 GL Nijmegen, The Netherlands\\
    \email{M.Janssen@astro.ru.nl}
   \and Max-Planck-Institut f\"ur Radioastronomie, Auf dem H\"ugel 69, D-53121 Bonn, Germany
            \and
            Steward Observatory and Department of Astronomy, University of Arizona, 933 N. Cherry Ave., Tucson, AZ 85721, USA
            \and
            Data Science Institute, University of Arizona, 1230 N. Cherry Ave., Tucson, AZ 85721, USA
            \and
            Program in Applied Mathematics, University of Arizona, 617 N. Santa Rita Ave., Tucson, AZ 85721, USA
            \and
            Department of Astrophysical Sciences, Peyton Hall, Princeton University, Princeton, NJ 08544, USA
            \and
            NASA Hubble Fellowship Program, Einstein Fellow
            \and
            Instituto de Astrofísica de Andalucía-CSIC, Glorieta de la Astronomía s/n, E-18008 Granada, Spain
             }

   \date{Received TBD; accepted TBD}

\nocite{eht-paperI}
\nocite{eht-paperIII}
\nocite{eht-paperVI}
\nocite{eht-m87-paper-viii}
\nocite{eht-SgrAi}
\nocite{eht-SgrAii}

\abstract
{In the first two papers of this publication series, we present a comprehensive library of synthetic Event Horizon Telescope (EHT) observations and used this library to train and validate Bayesian neural networks for the parameter inference of accreting supermassive black hole systems. The considered models are ray-traced general relativistic magnetohydrodynamic (GRMHD) simulations of \sgra and \m87.}
{In this work, we infer the best-fitting accretion and black hole parameters from 2017 EHT data and predict improvements that will come with future upgrades of the array.}
{Compared to previous EHT analyses, we considered a substantially larger synthetic data library and the most complete set of information from the observational data. We made use of the Bayesian nature of the trained neural networks and apply bootstrapping of known systematics in the observational data to obtain parameter posteriors.}
{
Within a wide GRMHD parameter space, we find \m87 to be best described by a spin between 0.5 and 0.94 with a retrograde MAD accretion flow and strong synchrotron emission from the jet. \sgra has a high spin of $\sim$\,0.8 -- 0.9 and a prograde accretion flow beyond the standard MAD/SANE models with a comparatively weak jet emission, seen at a $\sim$\,\ang{20} -- \ang{40} inclination and $\sim$\,\ang{106} -- \ang{137} position angle. While previous EHT analyses could rule out specific regions in the model parameter space considered here, we are able to obtain narrow parameter posteriors with our \z{} framework without being impacted by the unknown foreground Faraday screens and data calibration biases. We further demonstrate that the Africa Millimeter Telescope extension to the EHT will reduce parameter inference errors by a factor of three for non-Kerr models, enabling more robust tests of general relativity.
 }
{Our results agree with multiwavelength constraints from the literature. It will be instructive to produce new GRMHD models with the inferred interpolated parameters for in depth model-data comparisons and to study their accretion rate plus jet power.}


   \maketitle

\ActivateWarningFilters[hreflink]
\section{Introduction}

With the \ac{eht}, we imaged the supermassive black holes in the centers of Messier 87* \citep[\m87,][]{eht-paperI} and Sagittarius~A* \citep[\sgra,][]{eht-SgrAi}. In \citet{eht-paperVI, eht-SgrAv, eht-m87-paper-viii, 2023EHTStokesV, eht-SgrAviii}, selections of \ac{grmhd} models are scored against specific observational \ac{eht} and multiwavelength data products.

\begin{figure*}
    \centering
    \includegraphics[width=0.445\textwidth]{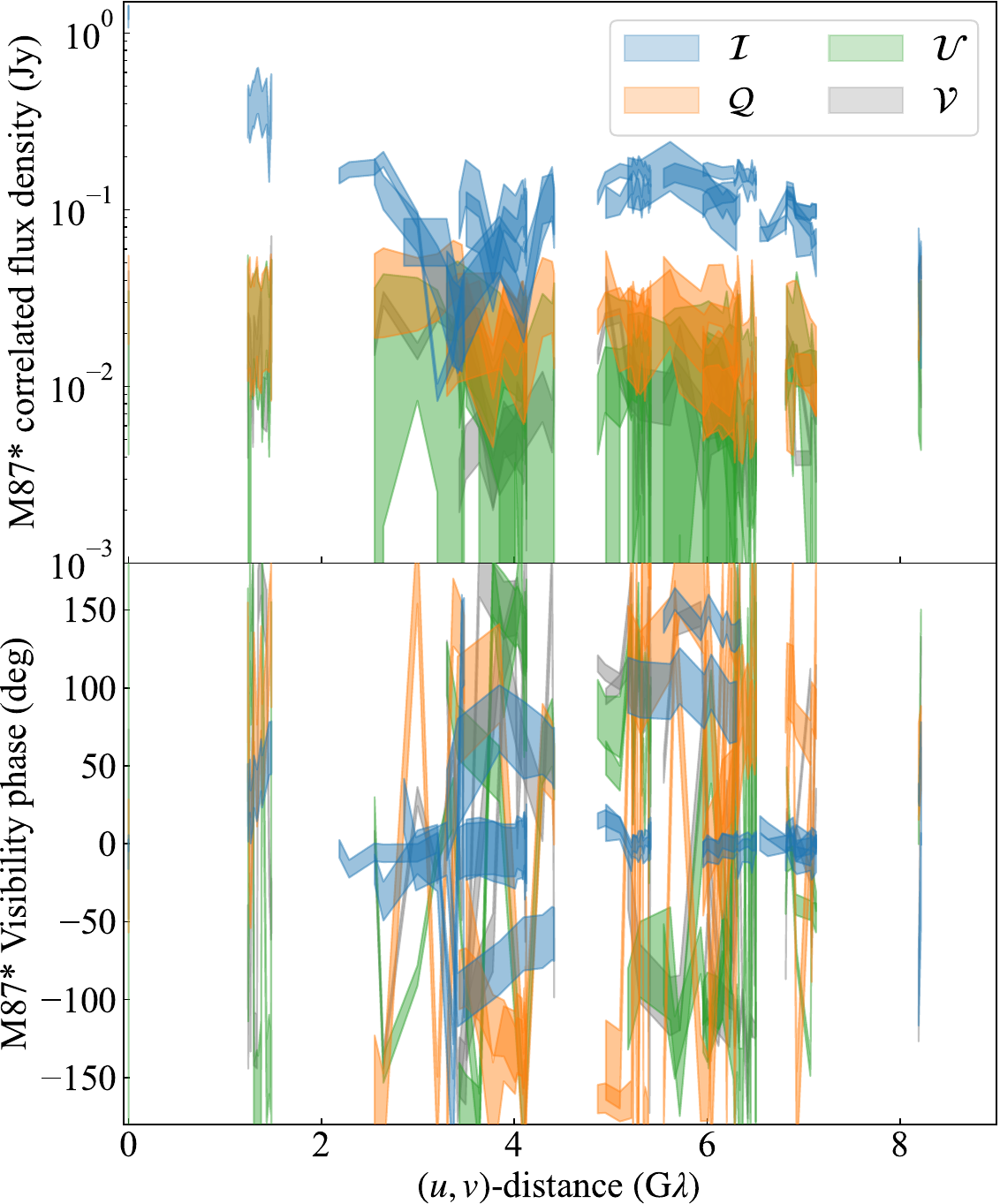}
    \includegraphics[width=0.445\textwidth]{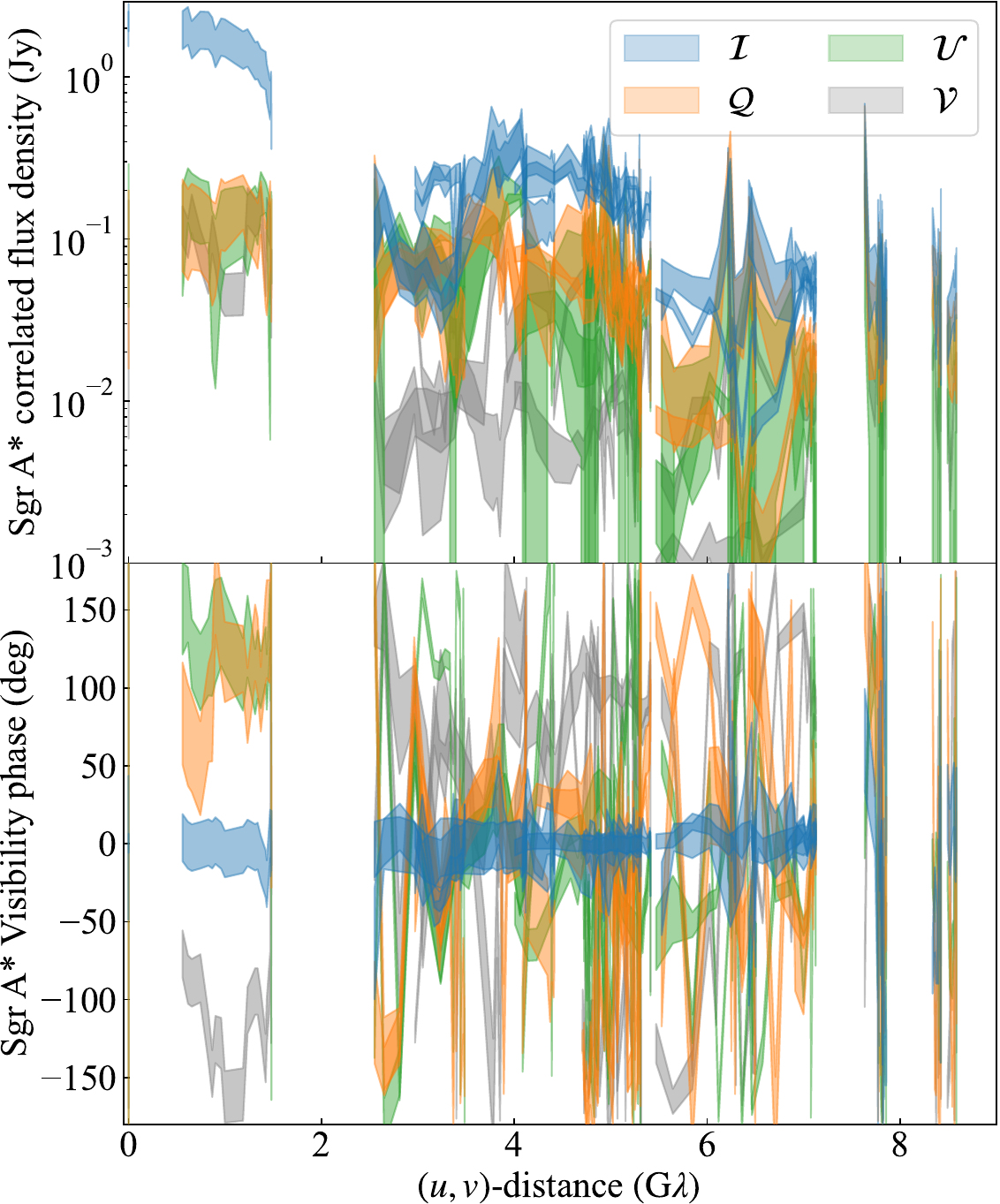}
    \caption{
       Correlated flux densities in Jansky (Jy) and visibility phases in degrees (deg) with standard deviation error bands computed from 1000 bootstrapping realization of the 2017 April 11 \m87 (left panels) and April 7 \sgra (right panels) observational \ac{eht} data. The measurements are plotted as a function of baseline length in units of the $\lambda\,1.3$\,mm observing wavelength.
       The displayed Stokes parameters show the total intensity ($\mathcal{I}$), linear polarization ($\mathcal{Q} \& \mathcal{U}$), and circular polarization ($\mathcal{V}$).
    } \label{fig:obsdata}
\end{figure*}

The scoring is facilitated through electromagnetic observables computed from the \ac{grmhd} models; the synchrotron emission at a wavelength of 1.3\,millimeter is predicted with general relativistic ray-tracing methods \citep[e.g.,][]{2020Gold, 2023Prather}.
Below, we list the key physical parameters of interest in the ray-traced \ac{grmhd} models considered in this study:
\begin{enumerate}
    \item Given the black hole mass $M$ and angular momentum $J$ measured with respect to the accretion flow, the dimensionless black hole spin $a_*$, given via $a_* = c J G^{-1}M^{-2}$, where $c$ and $G$ are the speed of light and the gravitational constant, respectively. For our ``standard'' models that we mostly considered here, the static Kerr spacetime metric \citep{Kerr_1963} varies only with $a_*$, as we fix the masses for \sgra and \m87 to be $4.14 \times 10^6\,M_\odot$ and $6.2 \times 10^9\,M_\odot$, respectively.
    \item For the considered Kerr-Newmann solutions, the dimensionless black hole charge $q_*$ \citep{Kerr-Newman_metric}. These models were ray-traced only in Stokes~$\mathcal{I}$.
    \item For the models that go beyond general relativity, the influence of a dilaton scalar field in the Einstein-Maxwell-Dilaton-Axion theory of gravity \citep{dilatonBH}. The dilaton \ac{grmhd} models used in this work are described in detail in \citet{Mizuno2018} and \citet{Roder2022IP}. These models were ray-traced only in Stokes~$\mathcal{I}$.
    \item The MAD or SANE magnetic state of the accretion disk \citep{eht-paperV, eht-SgrAv}. Most MAD models are more variable and launch more powerful jets compared to their SANE counterparts.
    \item The coupling between the temperatures of the protons and electrons $R_\mathrm{high}$ \citep{2016Moscibrodzka}. It is mostly the temperature of the electrons in the accretion disk and less so the strongly magnetized jet region that is sensitive to the $R_\mathrm{high}$ parameter.
    \item For \sgra, the orientation (or position angle) $\theta_\mathrm{PA}$ of the source on the sky and the inclination angle $i_\mathrm{los}$, with which the source is oriented with respect to our line of sight. Both $\theta_\mathrm{PA}$ and $i_\mathrm{los}$ parameters are measured relative to the accretion ﬂow angular momentum vector.
\end{enumerate}

In \citet{zingularity1}, a library of synthetic \ac{eht} observations of \m87 and \sgra are created from a library of \ac{grmhd} simulations that span the model parameter space listed above. Those parameters are then used as labels for the synthetic data to enable the supervised learning of Bayesian artificial neural networks (BANNs). We make use of the \z{} framework for the BANN training and validation, as described in \citet{zingularity2}. Here, we applied two trained networks to observational \sgra and \m87 \ac{eht} data.

Compared to previous \ac{eht} analyses, we utilized higher-quality observational data produced by an improved calibration scheme as described in \citet{zingularity1}. Furthermore, we considered a wider range of GRMHD models and systematics that affect the \ac{eht} data than prior works.
Our BANN implementations were trained to infer GRMHD model parameters based on salient features in the \m87 and \sgra \ac{eht} data with conservative uncertainty estimates \citep{zingularity2}.
For the \ac{grmhd} parameter inference, we considered the full Stokes information content of the \ac{eht} for the first time, instead of relying on derived data quantities.

In \Cref{sec:data} of this work, we describe the data used for the BANN training and parameter inference.
In \Cref{sec:nn}, we briefly review the training of our networks and then show the results when applying the trained BANNs to observational data in \Cref{sec:results}.
These results are discussed in \Cref{sec:discuss} and we present our conclusions in \Cref{sec:conclude}.
We finish with a description of the data and code availability in \Cref{sec:reproducibility}, allowing others to reproduce our results.

\begin{figure*}
    \centering\offinterlineskip
    \includegraphics[height=14.6cm, trim=0 0 0 0]{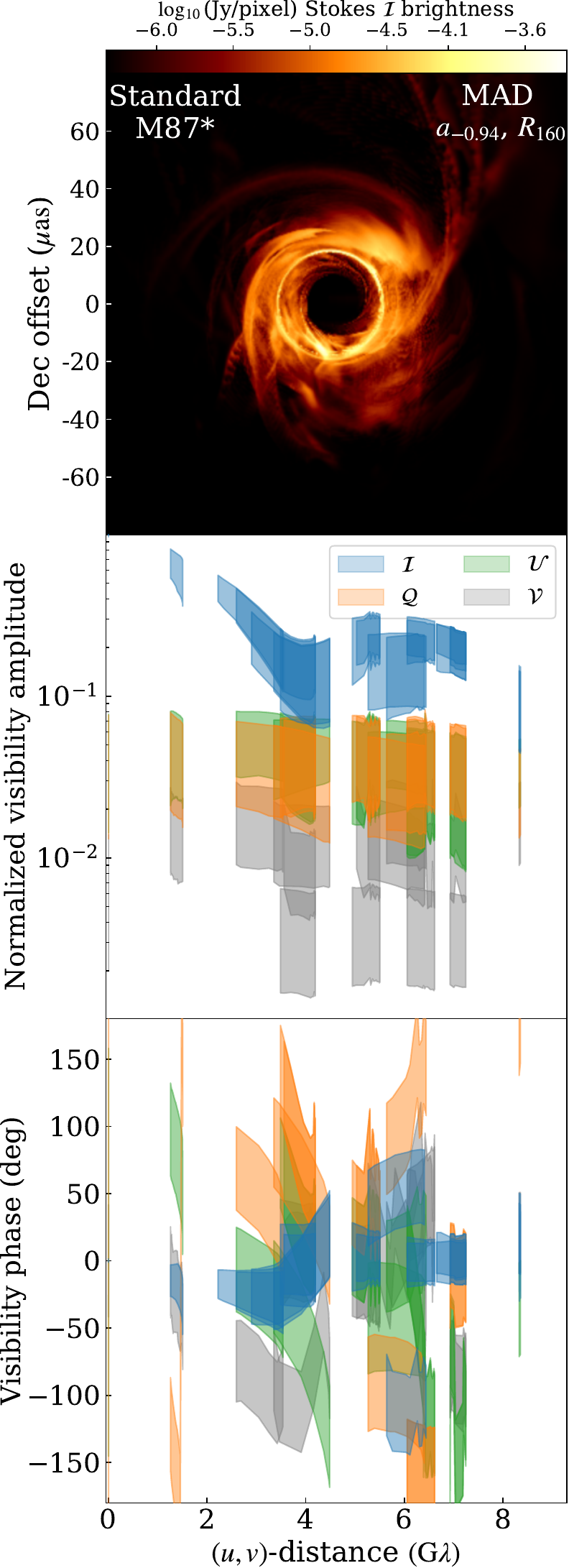}
    \hskip -0.6ex\includegraphics[height=14.6cm, trim=0 0 0 0]{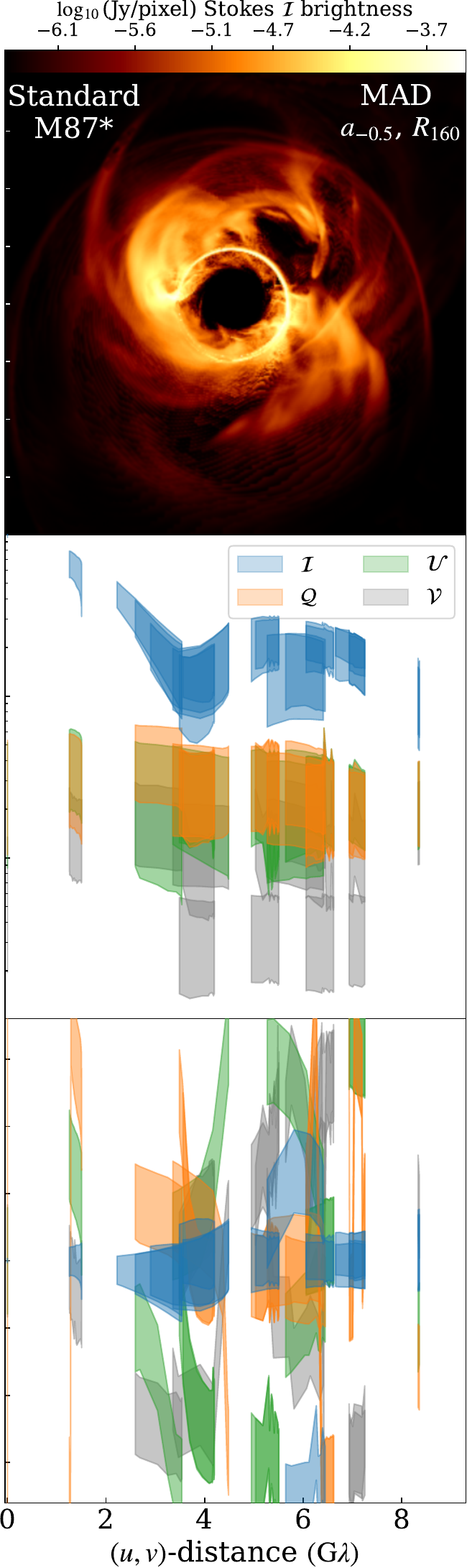}
    \hskip -0.6ex\includegraphics[height=14.6cm, trim=0 0 0 0]{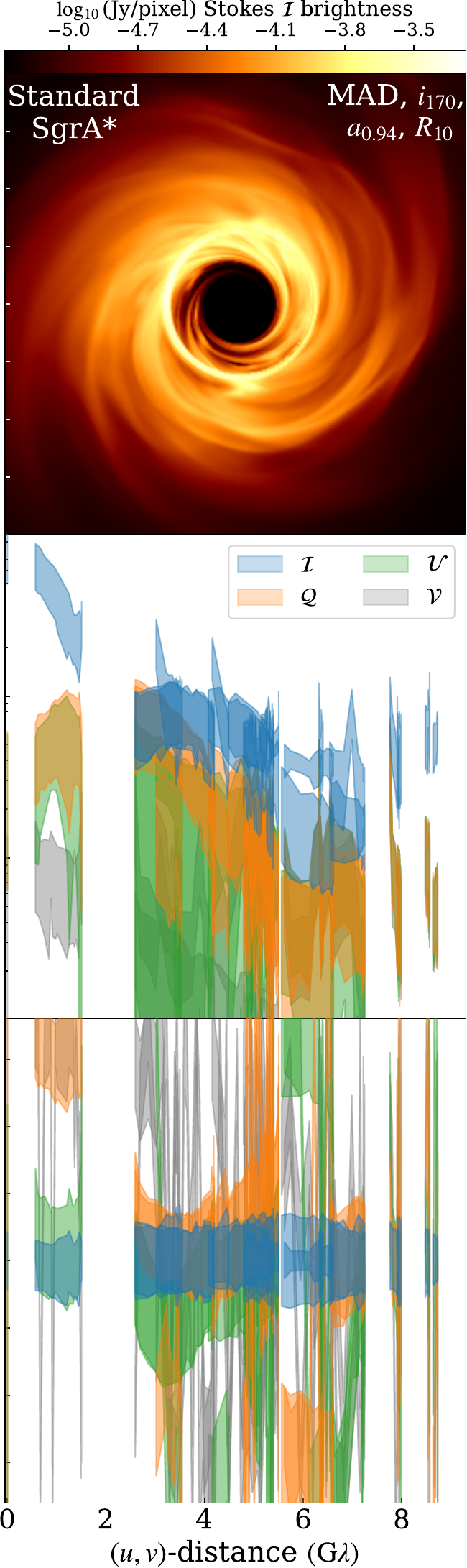}
    \hskip -0.6ex\includegraphics[height=14.6cm, trim=0 0 0 0]{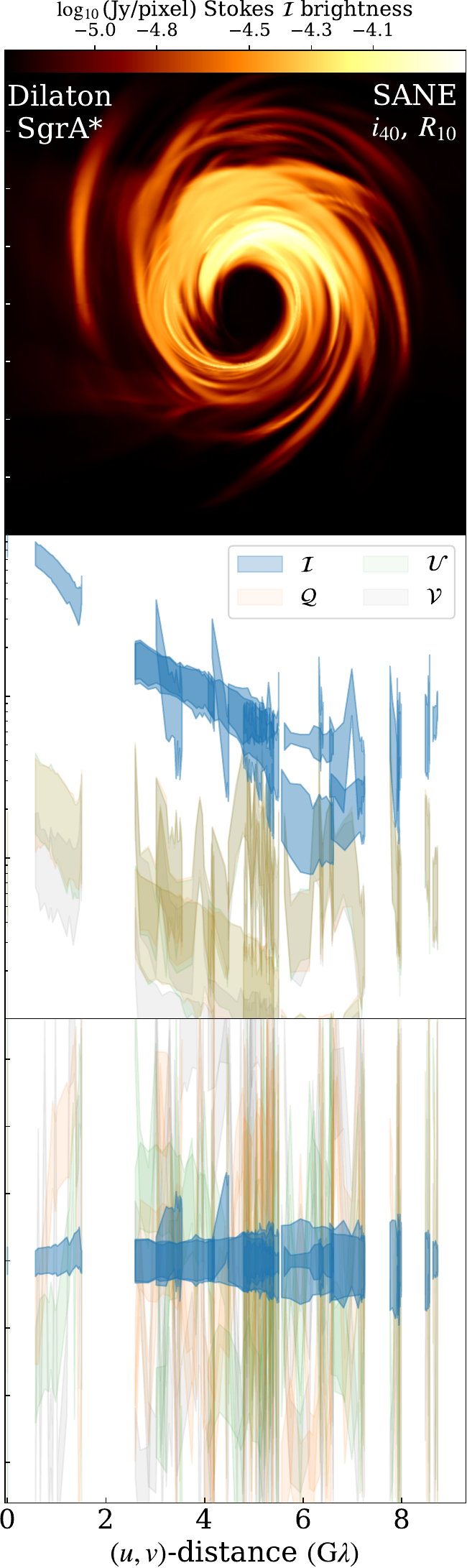}
    \caption{Visualization of training datasets with parameters close to our best-fitting BANN inferences. The top row displays example total intensity ray-traced ground-truth model images on logarithmic scales with varying dynamic ranges. Top left corners show the type of model, top right corners show the model parameters: spin $a_*=s$, $R_\mathrm{high}=r$, and $i_\mathrm{los}=l$ parameters are listed in a shorthand notation as $a_s$, $R_r$, and $i_l$. The \sgra models are displayed here with $\theta_\mathrm{PA}=0$. The standard models are based on the Kerr metric (see \Cref{sec:nn} text). Distributions of normalized visibility amplitudes and phases with standard deviation error bands computed from all synthetic data training samples of each model are plotted in the middle and bottom rows, respectively. These visibilities are depicted for all Stokes parameters. For the unpolarized dilaton models, the polarized data probes only instrumental effects and are faded out.}
    \label{fig:training_data}
\end{figure*}

\section{Data}
\label{sec:data}

The \ac{eht} interferometric array measures Fourier components of the sky brightness distribution at millimeter wavelengths. The coverage of the projected baseline vectors between pairs of \ac{eht} antennas is commonly given as $(u, v)$ vectors in units of the observing wavelength.

The observational data used in this work was taken by the \ac{eht} in 2017 \citep{eht-paperIII, eht-SgrAii} within a 226.1\,-\,228.1\,GHz frequency band.
We based our analysis on the $(u, v)$ coverage and data taken on April 7 for \sgra observations and April 11 for \m87.
The coverage and noise properties of the measurements were used for the neural network training with \ac{grmhd} synthetic data \citep{zingularity1}.
Amplitudes and phases of the observational data used for the subsequent \ac{grmhd} parameter inference are shown in \Cref{fig:obsdata}.

Synthetic datasets with parameters close to our BANN posteriors are shown in \Cref{fig:training_data}, alongside their corresponding ground-truth \ac{grmhd} models.
Here, we can see how only the linear polarization data shows distinguishing features between the two \m87 models with different spins.
We note that we generally do not expect an exact match between the \Cref{fig:obsdata} observational and \Cref{fig:training_data} synthetic data. Firstly, the \Cref{fig:training_data} models only roughly match the inferred optimal parameters by our BANNs; a limitation of the \ac{grmhd} sampling. We know that small changes in model parameters can have significant effects on the visibility data \citep[Section 7.3 of][]{zingularity2}. Secondly, the BANN inference does not use all of the data equally but focuses on salient features that can be used to differentiate between model parameters. For example, the Stokes~$\mathcal{U}$ phase of \sgra at a $(u, v)$-distance of around 1G$\lambda$, which is sensitive to the larger scale orientation of the polarization, does not match between the shown Kerr (``standard'') polarized \sgra model data and the observations.
Given the similarity of the model parameters, the Stokes~$\mathcal{I}$ visibilities from the Kerr and dilaton \sgra models are similar even though the underlying spacetime metrics are different.

We made use of observational `CASA' data calibrated with the \rpicard{} \citep{2019Janssenproc, Janssen2019} pipeline.
The characteristics of the BANN training sets are akin to the CASA data \citep{2020Roelofs}.
In \citet{zingularity1}, we presented a recent upgrade to the CASA data, which we make use of for the first time here, and describe the synthetic training data generation process in detail. 

\section{Neural network training}
\label{sec:nn}

\begin{figure*}
    \centering
    \includegraphics[width=0.33\textwidth]{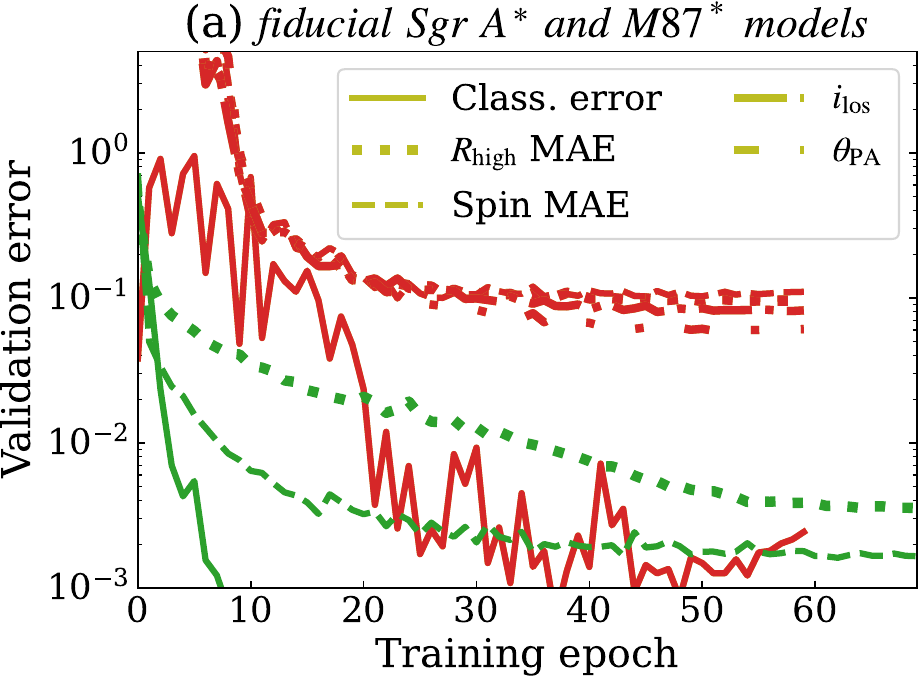}
    \includegraphics[width=0.33\textwidth]{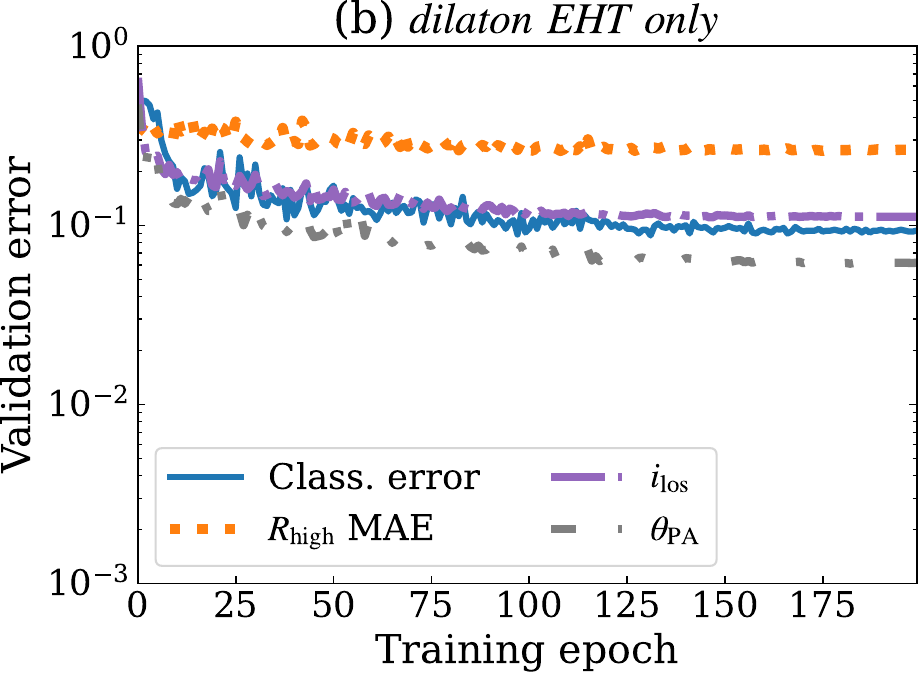}
    \includegraphics[width=0.33\textwidth]{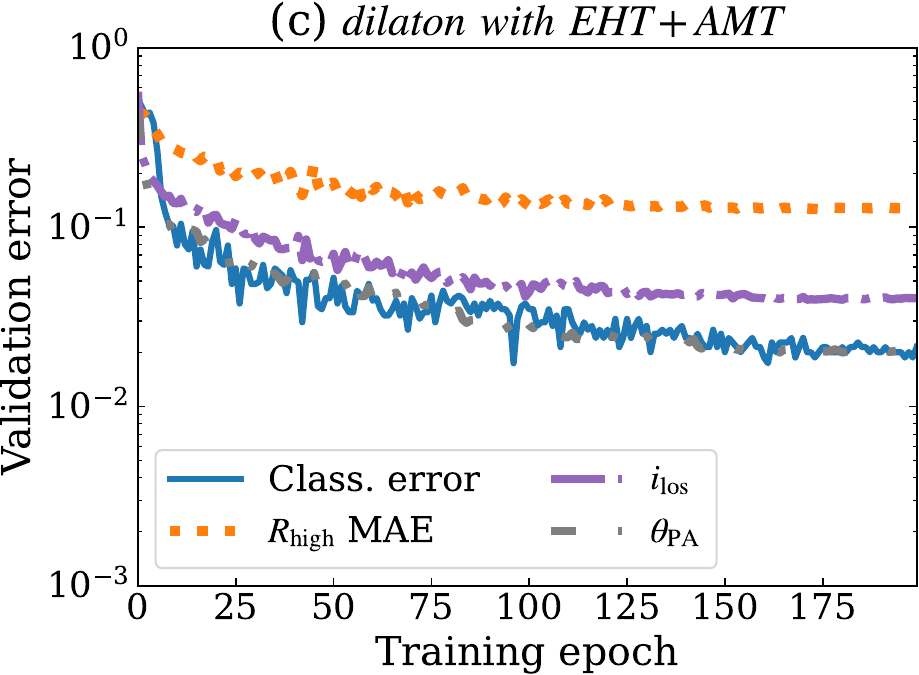}
    \includegraphics[width=0.33\textwidth]{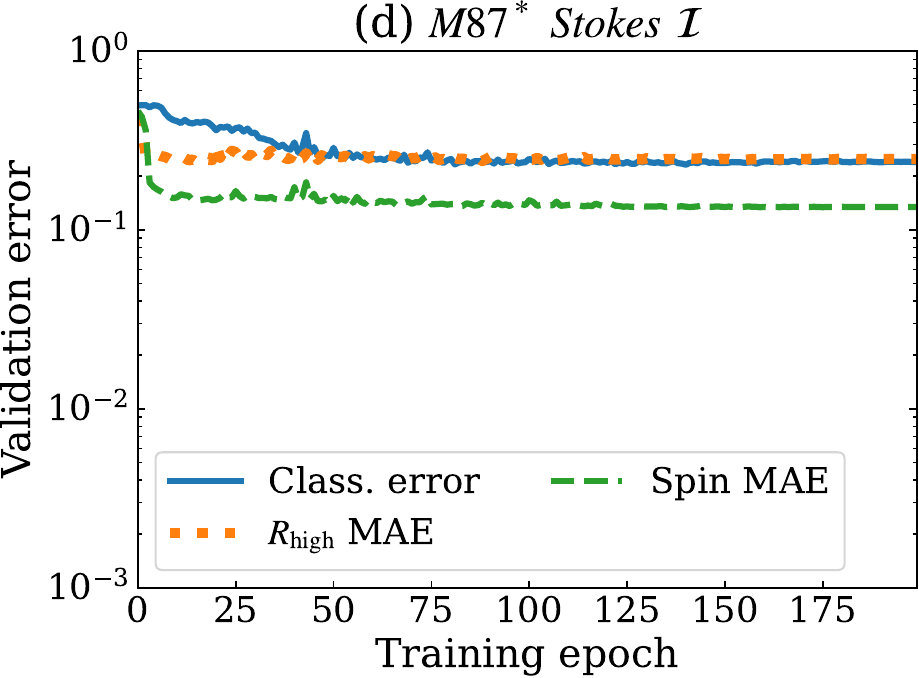}
    \includegraphics[width=0.33\textwidth]{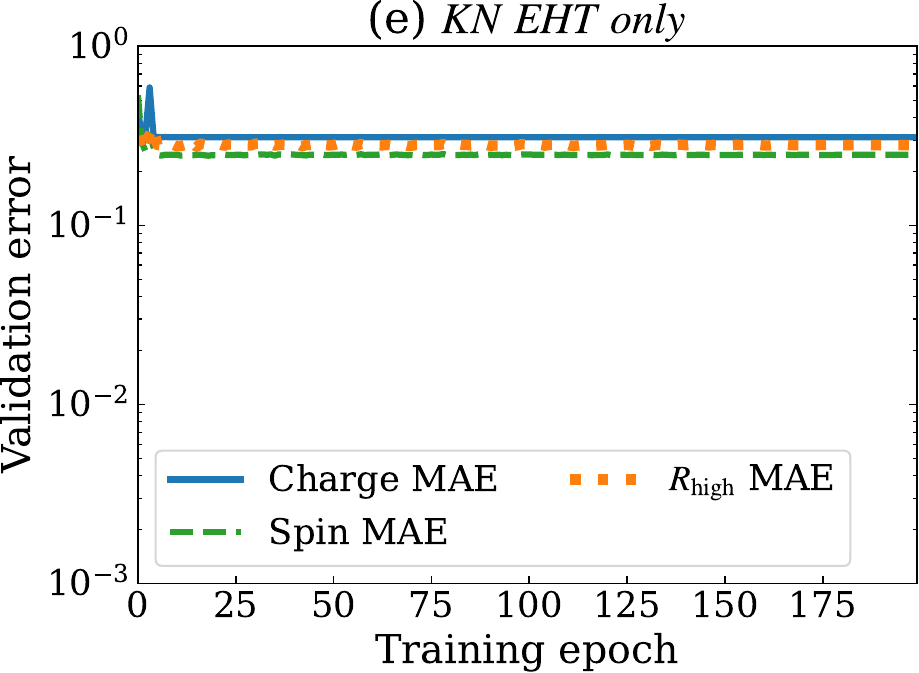}
    \includegraphics[width=0.33\textwidth]{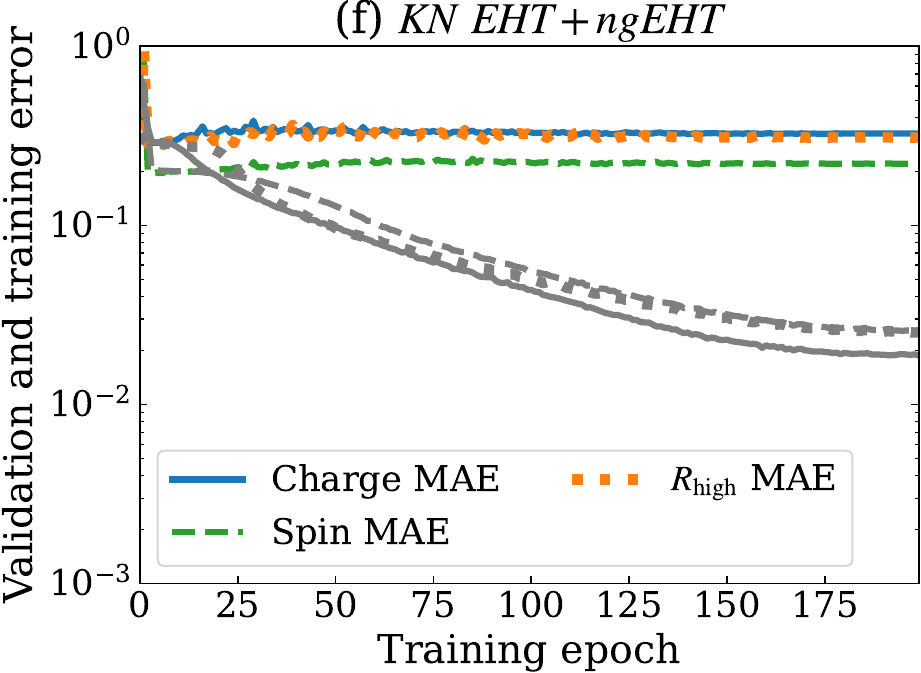}
    \caption{\z{} performance diagnostics for various neural network training runs. The validation error is computed from normalized labels of validation data not seen by the network during training. The mean absolute error (MAE) is computed as the average of all validation samples for normalized regression labels. The classification error (Class. error) is defined as one minus the network's accuracy, i.e., the fraction of misclassified validation samples. Panel (a) shows the fiducial models for \sgra and \m87 in red and green, respectively. Here, we show the \sgra model with a training length of 60 epochs; the 50 epoch model looks equivalent. For \m87, validation errors are overall smaller compared to \sgra and the classification errors get numerically close to zero beyond the logarithmic y-axis limit displayed in the figure here. Panel (f) also shows the training errors as gray curves.}
    \label{fig:zing_validation}
\end{figure*}

We used the \z{} framework to select the best BANN architectures for \sgra and \m87 from a survey and trained them on our \ac{grmhd} synthetic data library as described in \citet{zingularity2}. For Kerr \sgra, we two equally viable models, that only differ in the number of training epochs (60 vs. 50). For our other networks (\m87 and dilaton \sgra), we found single sets of best parameters that are used to make single fiducial models. 
For the training, we used the Swish activation function \citep{2017arXiv171005941R} and \mbox{RMSProp} optimization algorithm. For both \sgra and \m87, we employed similar network architectures, but with different numbers of connections and regularization methods. These hyperparameters were determined through parameter surveys. The employed Bayesian architecture is a combination of a ResNet \citep{ResNet} with subsequent variational fully connected layers.

The network training diagnostics of our fiducial models can be compared against our exotic models shown in the same \Cref{fig:zing_validation}. Similarly, we can compare the capabilities of the current \ac{eht} to predictions for planned array upgrades.
For our standard \sgra and \m87 BANNs, we achieved typical validation errors of about 10\,\% and 0.3\,\%, respectively.

The synthetic data generation process and BANN architectures used for the training are the same for the standard and exotic models.
As the exotic models were ray-traced only in total intensity, the training is limited to Stokes~$\mathcal{I}$ data, leading to relatively poor performance.
For the dilaton models, we see a similar network performance compared to the \sgra Stokes~$\mathcal{I}$ test of \citet{zingularity2}.
A comparison of panel (b) with (c) shows how the planned Africa Millimeter Telescope \citep[AMT,][]{2016AMT} improves validation errors of the dilaton model parameters by a factor of three on average. For a fair comparison, we used the same 2017 EHT antennas with identical sensitivities in both cases.
We attribute the strong impact of the AMT primarily to the increased northeast and southwest resolution \citep{2023LaBella} as well as the close $(u, v)$ baselines from the South Pole to Chile and the AMT. These ``crossing'' tracks can be used to remove systematics from the data after the close-together Chile sites have been calibrated based on the total flux density of the source \citep[e.g., Section 4.2 in][]{2022Janssen}.

The bottom panels of \Cref{fig:zing_validation} present our network training tests for Kerr-Newman models. As a baseline test, we first train on our standard \sgra models when only using total intensity data: Validation errors close to 10\,\% are reached only for the spin parameter, as shown in panel (d).
For the Kerr-Newman models, the network failed to train on the synthetic data and no reliable estimations could be obtained for any parameter, as shown in panel (e). We attribute the failure of fitting spin to the degeneracy with charge in the Kerr-Newman metric \citep{zingularity1}.
With additional dishes from the ngEHT project \citep{2019ngEHT, 2023ngEHT} joining EHT observations, only marginal improvements can be achieved, as shown in panel (f).
The errors on the training data itself follow the validation errors in (a) to (e).
For (f), the training error differs from the validation error and is decreasing. With the amount of data from the extended baseline coverage, it will likely be possible to find a suitable network architecture to train on the Kerr-Newman features without the problematic overfitting seen here.

\section{GRMHD parameter inference}
\label{sec:results}

For each observational dataset, we created 1000 bootstrapped realizations with varying polarization leakage, partially correlated telescope gain plus gain curve errors, and thermal noise as described in Section 6.2 of \citet{zingularity2}.
Parameter posteriors were formed from 1000 inference passes through our BANNs for each realization of the observational data.

\begin{figure}
    \centering
    \includegraphics[width=\columnwidth]{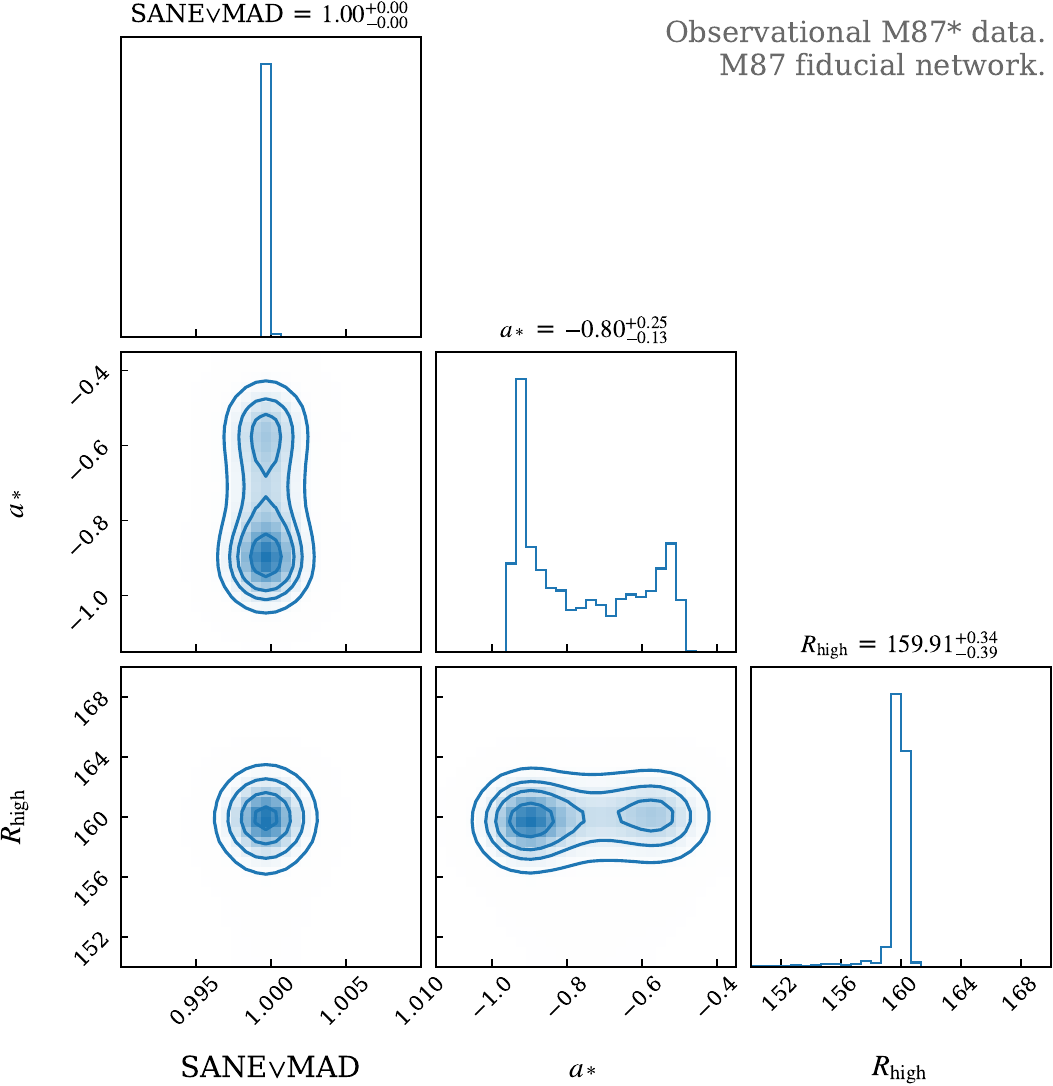}
    \caption{Posterior obtained from $10^6$ variational inference draws of 1000 bootstrapped realizations of the observational \m87 data with our fiducial \m87 BANN. For the magnetic state, a value of zero corresponds to a certain SANE classification, and a value of one to a certain MAD classification.}
    \label{fig:M87fiducial_e17e11loOBS}
\end{figure}

\Cref{fig:M87fiducial_e17e11loOBS} shows the results from applying our fiducial \m87 BANN to the April 11 observational data.
The data favor MAD models with a large $R_\mathrm{high} = 160$ parameter. These models produce powerful outflows with a strong jet contribution to the synchrotron emission. The training data had a maximum $R_\mathrm{high}$ value of 160, so it might be that a model with an even higher $R_\mathrm{high}$ would describe the data better. The equivalent high lepton energies could be produced by nonthermal effects such as magnetic reconnection, which would give a different energy distribution function. 
For the spin, our network infers values in the range of -0.5 to -0.94, clearly preferring retrograde accretion.

\begin{figure}
    \centering
    \includegraphics[width=\columnwidth]{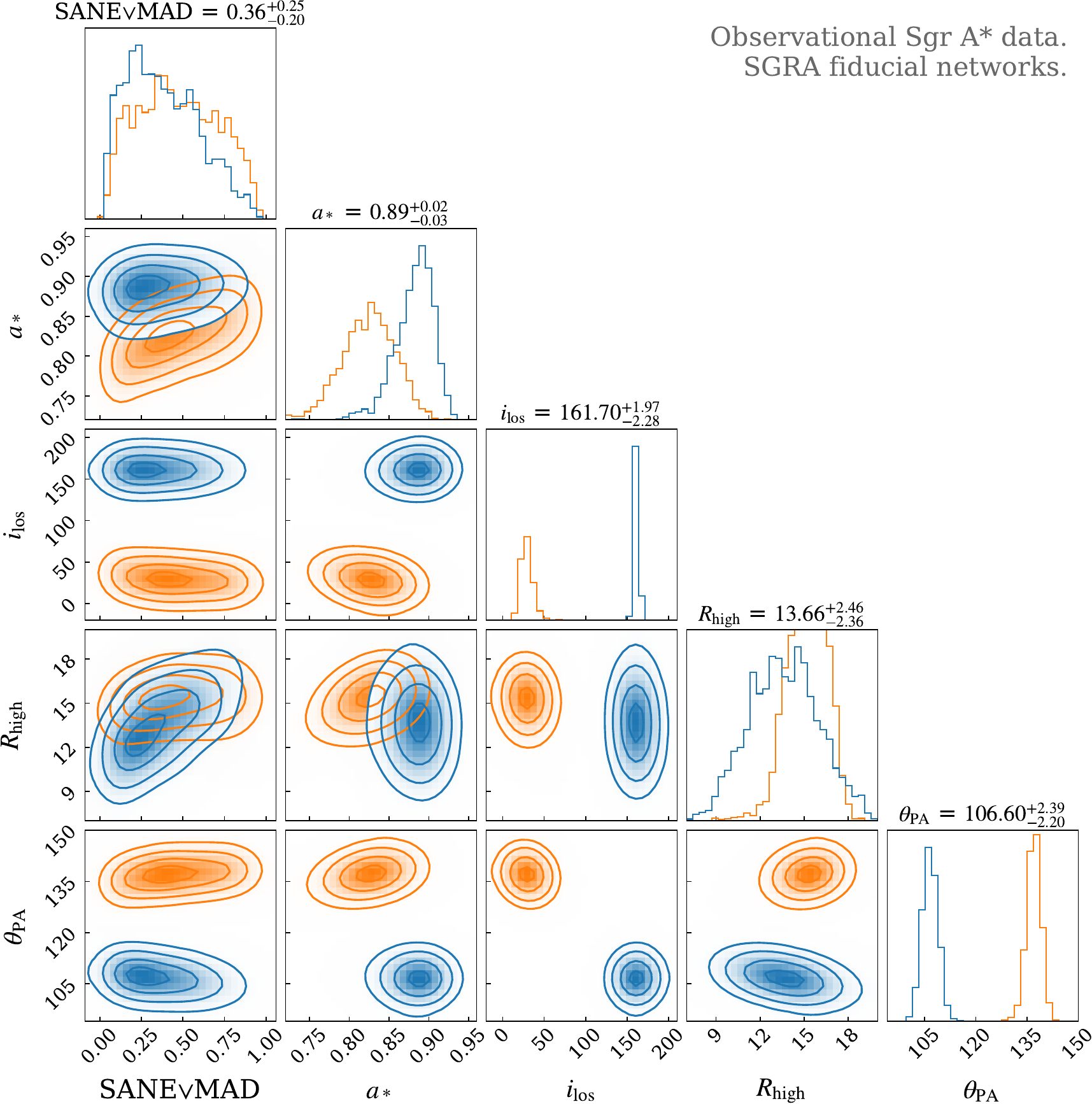}
    \caption{Same as \Cref{fig:M87fiducial_e17e11loOBS} but for \sgra data using our two fiducial \sgra BANNs for which the posteriors are shown in blue and orange, respectively. Based on measurements from other instruments (see text), we have a slight preference for the model that predicts $i_\mathrm{los} > \ang{90}$ (shown in blue), for which we give the numerical parameter ranges in the figure.}
    \label{fig:SGRAfiducial_e17c07loOBS}
\end{figure}

\begin{figure}
    \centering
    \includegraphics[width=\columnwidth]{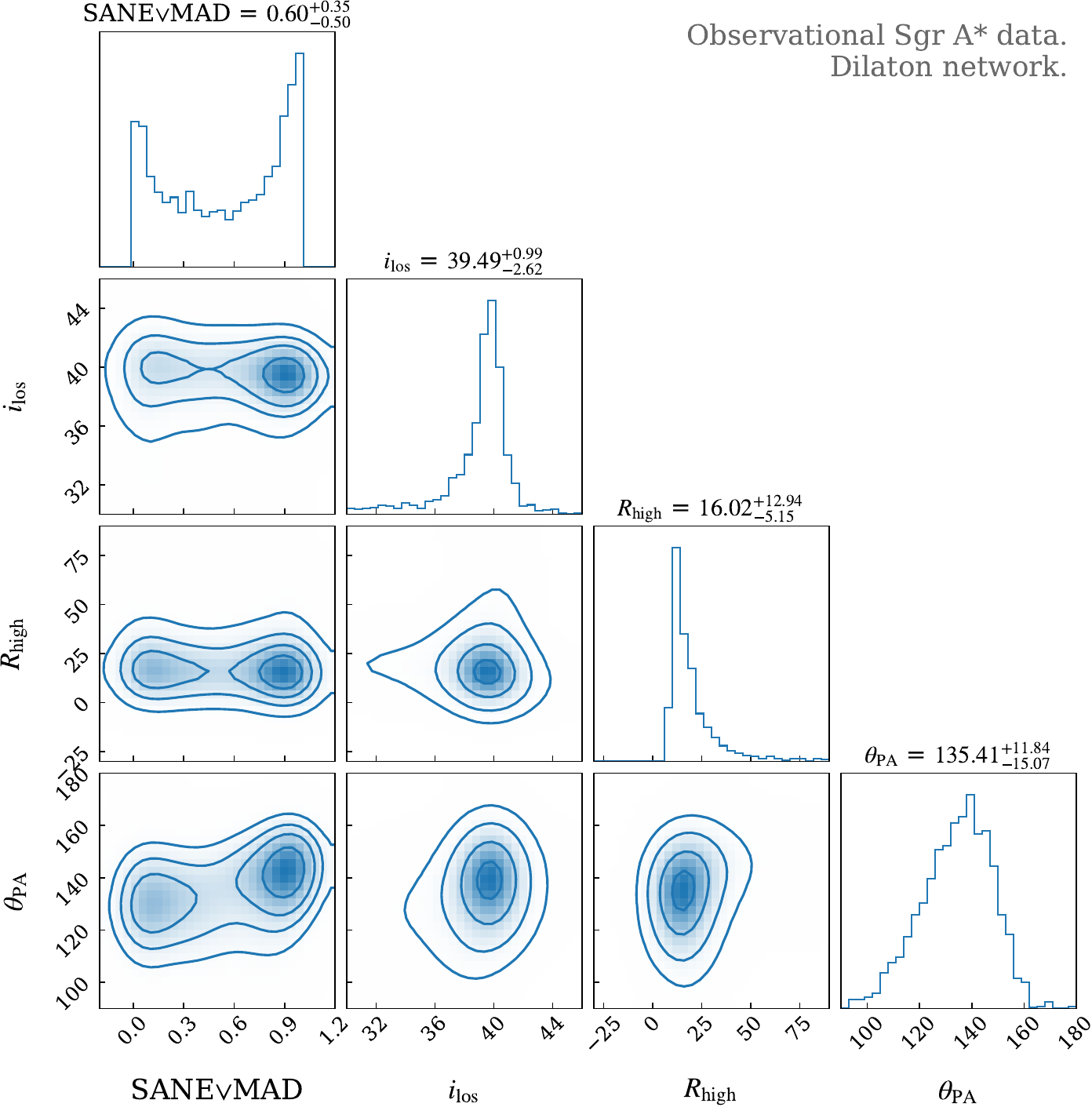}
    \caption{Same as \Cref{fig:M87fiducial_e17e11loOBS} but for \sgra data using our dilaton BANN. The models underlying the dilaton training data were only ray-traced for $i_\mathrm{los} < \ang{90}$ values.}
    \label{fig:Dilaton_e17c07_lo_SGRA}
\end{figure}

\Cref{fig:SGRAfiducial_e17c07loOBS} shows the results from applying our two fiducial \sgra BANNs to the April 7 observational data.
The data are inconclusive concerning the magnetic state of the accretion flow. MAD models describe the polarization quantities well, while SANE models are less problematic in terms of excess variability compared to the observations, but see also \citet{2024Salas}. A model beyond the standard MAD/SANE dichotomy might work better here.
The spin parameter gives a clear preference toward high $\sim0.8$ -- 0.9 values and a prograde accretion flow.
Furthermore, the spin axis is oriented close to the line of sight at an angle of about \ang{162} (\ang{29} for the other model) and at $\theta_\mathrm{PA} \sim \ang{106}$ -- \ang{137} east of north in the plane of the sky. Due to the symmetry of the GRMHD models, \ang{162} $i_\mathrm{los}$ corresponds to \ang{18} but for an opposite sense of rotation of the accretion flow. Within the uncertainty from our $i_\mathrm{los}$ training data sampling in \ang{20} steps, the two BANNs consistently predict small inclination angles of \sgra's spin axis with respect to our line of sight. Evidently, the direction of the accretion flow direction cannot be discerned.
The difference between the two position angles inferences can also be understood from the $\theta_\mathrm{PA} = \ang{90}, \ang{120}, \ang{150}$ sampling of the training data: the inferred values fall on opposite sites of the middle grid value.
The low $R_\mathrm{high}\sim14$ value corresponds to a dominant emission from the accretion flow via hot electrons in the disk. 

These \sgra results have been obtained with the level of data calibration described in \citet{eht-SgrAii} and \citet{zingularity1}. We note that consistent posteriors with no visible differences are obtained from data that has been processed with additional model-dependent calibration steps described in Section 2.2 of \citet{eht-SgrAiii}. We attribute this to the robustness of our BANNs against gain errors in the data, learned from the simulation of data corruption effects in the training data.
The \citet{eht-SgrAviii} GRMHD scoring results depend on assumptions made about the origin of the Faraday rotation measure (RM). The RM has a constant external component and an internal time-variable component \citep{2021Goddi, 2024Wielgus}. Only the internal component can be simulated in the small computational domain of our GRMHD models. We found that our BANN is not using the constant RM component as a salient model-discriminating data feature; we get identical posteriors when de-rotating the \sgra Stokes~$\mathcal{Q, U}$ visibilities by 50\,\% or 100\,\% of the overall measured $\mathrm{RM} = -4.65 \times 10^5\,\mathrm{rad}/\mathrm{m}^2$. Applying a time-variable RM to the data on the other hand does change the results. Evidently, our BANN has trained on the internal time-variability of $\mathcal{Q, U}$ phases as discriminating factors for $a_*$ and $i_\mathrm{los}$.
We note that the images across our model library do show differences in their overall electric vector position angle orientations, which is equivalent to having different constant RM components.

\Cref{fig:Dilaton_e17c07_lo_SGRA} is the posterior from the April 7 observational data obtained from our dilaton BANN.
All parameters, in particular $R_\mathrm{high}\sim16$ and $\theta_\mathrm{PA} \sim \ang{135}$, agree well with the standard model results.
For this model that goes beyond standard GR, also the MAD/SANE classification is inconclusive. We note that the dilaton model considered here is nonrotating and does not reach fully developed MAD states as the amount of magnetic flux accumulating near the event horizon remains modest. The small \ang{10} difference in inclination angle is well within the uncertainties and likely related to the different discrete $i_\mathrm{los}$ sampling values of the training data. Closest to the inferred values of \ang{29} and \ang{39}, these are $i_\mathrm{los} = \ang{10}, \ang{30}$ for the standard models and $i_\mathrm{los} = \ang{20}, \ang{40}$ for the dilaton models, respectively. We note that no $i_\mathrm{los} > \ang{90}$ mode can be inferred for the dilaton models, as the ray-tracing was only done up to a maximum inclination of \ang{60}.

The consistency of parameter inference between the dilaton and fiducial \sgra models is noteworthy, given that the dilaton BANN trained only on Stokes~$\mathcal{I}$ data, which we have shown to not work for \m87.
We attribute this to the intrinsic dilaton model variability measured across different baselines over the course of a single \ac{vlbi} observation. With the help of this variability, models with different parameters can be distinguished.

Our individual BANNs are interpolating between training values for many parameters\footnote{We impose no bound to the parameters our BANNs could infer; in fact it is easy to devise a network that would predict unphysical \ac{grmhd} model parameters.}, with no wide multimodal posteriors.
This latter case is a ``failure mode'' of our BANNs, when the variational inference runs on data that are difficult to characterize \citep{zingularity2}.
Compared to GRMHD validation data, wider posteriors are obtained on the observational data. Evidently, the observational data cannot be perfectly described by GRMHD models due to missing physics and/or the incomplete sampling of model parameters. Thus, the observational parameter uncertainties are larger than the BANN validation errors. Yet, GRMHD simulations are currently the best models we have to describe the horizon-scale emission of low-luminosity AGN in a self-consistent manner. 
Barring the SANE/MAD magnetic state inference, \sgra has relatively narrow posteriors even though the training validation errors are larger than for \m87. Without strong jet emission from accelerated particles and with an extremely low accretion rate, the \sgra data are quite well described by the ideal GRMHD models.

\section{Discussion}
\label{sec:discuss}

The GRMHD models underlying our BANN training data assume the presence of a magnetized turbulent accretion flow surrounding a supermassive compact object. The initial gaseous torus is aligned with the equatorial plane of the black hole (i.e., the accretion disk is not tilted).
The gas is assumed to be pure hydrogen. Pair production and radiative cooling processes are not simulated. 
The electron distribution function is assumed to be thermal, neglecting particle acceleration processes and nonideal MHD processes.
Even though our analysis is model-dependent, we sampled a broad parameter space. Thus, it is instructive to put our inference results into perspective.

Our \m87 analysis favors MAD models with large $R_\mathrm{high}$ values and $a_*$ between $-0.5$ and $-0.94$, which have powerful outflows and strong synchrotron emission from the jet. The probability density likely peaks at those particular values because they are two neighboring values in our GRMHD training data grid space. We also note that past EHT image scoring analyses have disfavored the MAD $a_* = -0.94$ models \citep{eht-paperV, eht-m87-paper-viii}. 
We thus argue for the ``true'' spin of a best-fitting GRMHD model to be most likely at an intermediate $-0.94 < a_* < -0.5$ value.
The strong preference for a MAD accretion in \m87 is in agreement with past EHT analyses.
Furthermore, the inferred model parameters satisfy jet power constraints measured at larger spatial scales \citep{eht-paperV, 2019Nemmen}.
The black hole counter rotation fits in the picture of \m87 being an elliptical galaxy likely affected by past mergers \citep[e.g.,][]{2007Volonteri, 2023Raimundo}. Mergers can naturally explain the large spin \citep[e.g.,][]{2008Berti}, while the black hole spin-down will not be too extreme in the strongly sub-Eddington accretion of \m87, even for a retrograde MAD flow \citep[e.g.,][]{2022Narayan, 2024Lowell}.

\citet{2023Qiu} trained a random forest machine learning model on a few predetermined EHT observables and infer high-spin retrograde models with large $R_\mathrm{high}$ values for \m87. Our findings are in excellent agreement with Qiu et al., who have used a similar GRMHD library but without the full forward modeling. 
In our work, we only considered models where the magnetic field polarity is aligned with the accretion disk angular momentum vector on large scales. \citet{2023Qiu} and \citet{2024Joshi} have shown that these models are preferred over models with anti-aligned polarity. The field polarity mainly affects circular polarization, which does not yield strong model constraints in the 2017 EHT data \citep{2021Ricarte, 2023EHTStokesV}.

The high spin value of $a_* \sim 0.8$ -- 0.9 inferred for \sgra by our fiducial network agrees with mounting evidence from independent analyses in the literature, which suggest a spin $> 0.5$.
From the \citet{2018Eckart} literature overview of several model-dependent radio, near-infrared, and X-ray data analyses, a slight tendency for $a_* > 0.5$ crystallizes.
Recent combined Chandra plus VLA modeling results find $a_* = 0.9 \pm 0.06$ when assuming the presence of a collimated outflow \citep{2024Daly}.
From the analysis of ALMA light curves, \citet{2022Wielgus} find hints of a positive spin and \citet{2024Yfantis} find $a_* >0.8$, but noted a weak spin-dependence of their results.
In early fits of GRMHD models to the \sgra spectral energy distribution (SED) and initial \ac{vlbi} size constraints, high spin values were also preferred \citep{2009Moscibrodzka}.
Finally, the best-bet model from our recent EHT analysis has $a_* = 0.94$ \citep{eht-SgrAviii}. One should note the fundamental difference of constraint-based EHT modeling compared to our BANN inference. Additionally, the EHT modeling assumed the RM to be external and discarded the model variability problem \citep{eht-SgrAv,Wielgus2022LC}. We attribute the stark difference in the electron temperature coupling factor -- $R_\mathrm{high} =160$ of the $a_* = 0.94$ model versus a much lower value inferred here -- to these fundamental analysis differences. Similarly, the discrepancy between the prograde accretion flow (positive $a_*$) inferred here and the preference for a retrograde MAD flow in \citet{2024Joshi} can be traced to the difference of our approach from the standard GRMHD scoring and the different observational data being used.
For a detailed discussion of how the black hole spin can be inferred from EHT data, we refer to \citet{2023Ricarte}.

We find \sgra's spin axis to be closely aligned with our line of sight, consistent with earlier findings based on GRAVITY polarimetric plus astrometric measurements of flares \citep{2018Gravity, 2020GRAVITY, 2023GRAVITY} as well as ALMA light curves \citep{2022Wielgus, 2024Levis, 2024Yfantis}. These studied $\mathcal{Q}-\mathcal{U}$ loops display a clockwise rotation on the sky.
For our two BANNs, the inferred values are $i_\mathrm{los} = 28.9^{+7.5}_{-6.3}$ degrees and $i_\mathrm{los} = 161.7^{+2.0}_{-2.3}$ ($18.3^{+2.3}_{-2.0}$) degrees. Taking the $\mathcal{Q}-\mathcal{U}$ loop measurements into account, $i_\mathrm{los} = \ang{162}$ is preferred, as inclinations larger than 90 degrees correspond to a clockwise accretion flow rotation on the sky in our GRMHD models.
Yet, it is worth noting that the super- (rather than sub-)Keplerian motion of a hot-spot describing the GRAVITY $\mathcal{Q}-\mathcal{U}$ loops \citep{2020Matsumoto, 2024Yfantisb} speaks for an emitting region outside of the standard GRMHD accretion flow. \citet{2020Matsumoto}, \citet{2023Lin}, and \citet{2024Antonopoulou} for example successfully fit the polarization loops with outflow models. For prograde accretion, it is expected that hot spots in outflows and/or current sheets outside of the bulk accretion flow would follow the accretion direction of rotation. Thus, our disfavored $i_\mathrm{los} = 28.9^{+7.5}_{-6.3}$ solution would require wind-fed accretion \citep[e.g.,][]{2023Ressler} scenarios, where the disk angular momentum can change on short timescales. In this case, counterclockwise $\mathcal{Q}-\mathcal{U}$ loops would be observed sometimes. In an alternative scenario, flux tubes in a counterclockwise accretion flow may be bent in the opposite direction, leading to entrapped hot spots moving clockwise \citep{2025Antonopoulou}.
For the accretion flow model considered in \citet{2024Faggert}, see also \citet{2022Ozel} and \citet{2023Younsi}, $a_* = 0.8$ agrees with the observed \sgra image brightness asymmetry for a broad range of sub-Keplerian velocity profiles for $i_\mathrm{los} = \ang{18}$, while $i_\mathrm{los} = \ang{29}$ requires a very slow angular rotation.
Generally, hot accretion flows are expected to be sub-Keplerian due to pressure gradients.

As we find \sgra unlikely to have a MAD accretion disk with a powerful outflow, the direction of a potential large-scale jet would be determined by the interstellar medium and not the black hole spin direction \citep[e.g.,][]{hamr1, 2023Kwan, 2023Ressler}.
Yet, \citet{2024Wang} show that a past merger with Gaia-Enceladus \citep{2018Helmi} can reproduce a high $a_*$ in \sgra with a low $i_\mathrm{los}$, where the BH spin axis is misaligned with the Milky Way's rotation.
Generally, a low $i_\mathrm{los}$ does thus not preclude \sgra jet activity to be responsible for the Galactic Fermi/eROSITA bubbles \citep[][note in particular the discussion in Section 4]{2024Sarkar}.
Recently, \citet{2023Ressler} and \citet{2024Galishnikova} have shown that the presence of strong magnetic fields in \sgra \citep{eht-m87-paper-vii} does not necessarily lead to a MAD state with strong jets.

The inferred $\theta_\mathrm{PA} \sim \ang{106}$ -- \ang{137} position angle of \sgra's spin axis matches with the $\sim \ang{135}$ found by \citet{2021Ball} and $\ang{130} \pm \ang{20}$ from \citet{2024Yfantisb} from the modeling of GRAVITY flares.
On the other hand, the \citet{2023GRAVITY} find $\sim \ang{177}\pm\ang{24}$ from the GRAVITY data, which is however clearly disfavored by \citet{2024Yfantisb}.
It is not yet understood why the modeling of polarimetric ALMA light curve data yields significantly different PAs in the range of \ang{0} to \ang{57} \citep{2022Wielgus, 2024Yfantis}, but note the \ang{180} degeneracy without the astrometry in the ALMA data (Diogo Ribeiro, priv. comm.). Possibly, ALMA measures an emission region from an inflow or outflow farther away, where the PA is not directly related to the black-hole spin axis.\footnote{As discussed in \citet{eht-SgrAv}, the question whether ALMA measurements are impacted by a slowly varying extended unresolved structure in \sgra is important for the reconciliation of the variability discrepancy between GRMHD models and ALMA light curves.}
As noted by Ball et al., $\theta_\mathrm{PA}$ close to \ang{150} are consistent with the angular momentum axis of a speculative cold disk around \sgra \citep{2019Murchikova}.
It thus seems more likely that the Doppler-shifted H30\,$\alpha$ ALMA measurements are indeed signatures of an accretion flow rather than a collimated jet as suggested by \citet{2019Royster} and \citet{2020YusefZadeh}. Although, as laid out earlier, a jet in \sgra may very well not follow the black hole spin direction on large scales.
Relatedly, we note that \sgra image asymmetries determined in current 3\,mm VLBI studies are not necessarily related to a jet direction. As noted by \citet{Issaoun_2019}, the intrinsic source structure is highly symmetric when de-scattering methods are applied and the image major axis direction may also follow accretion flow emission.
Future mm-VLBI upgrades will likely enable the \sgra jet detection through direct imaging \citep{2024Chavez}.

$R_\mathrm{high} \sim 14$ translates into the presence of relatively hot electrons in the \sgra accretion flow, which is in agreement with past SED model-fits \citep{2002Yuan, 2013Moscibrodzka}. 
As noted by {Mo{\'s}cibrodzka} and Falcke, $R_\mathrm{high}$ values between 10 and 30 are also predicted by viscous heating processes in shearing box simulations \citep{2007Sharma}. The MAD models with a magnetic turbulent cascade heating of electrons considered by \citet{2024Moscibrodzka} are most similar to $R_\mathrm{high} = 10$ images from the standard models in terms of observational characteristics.
The fact that no direct detection of a jet from \sgra has been made yet, fits into the picture of a source model with low $R_\mathrm{high}$, small $i_\mathrm{los}$, and no fully developed MAD accretion flow.

\section{Conclusions}
\label{sec:conclude}

We produced a comprehensive \ac{grmhd} synthetic data library based on the 2017 \ac{eht} data, reduced using an improved calibration pipeline. For the first time, we applied a Bayesian artificial neural network trained on our \ac{grmhd} library to \ac{eht} visibility measurements.
Combined with the bootstrapping resampling method, we used our \z{} networks to perform \ac{grmhd} parameter inference with a robust uncertainty estimation.

Through additional validation tests described here and in \citet{zingularity2}, we found that our networks have a persistent performance: The results are not influenced by noisy data features or the particular choice of (reasonable) network \mbox{(hyper-)parameters}. The strong intrinsic \ac{grmhd} model variability is also properly taken into account. 
By construction, our results are neither affected by assumptions made in the \ac{vlbi} data reduction strategy, nor the external physical influences of the \sgra scattering screen and external RM component.
We find our networks to be able to interpolate between the discrete \ac{grmhd} parameters from our training data.
It will be worthwhile to produce new \ac{grmhd} runs with these intermediate parameter values for future in-depth model-data comparisons. Additionally, the predicted accretion rate, jet power, and broadband spectral energy distribution can be studied.

For the \m87 data, we infer MAD models with $R_\mathrm{high} \geq 160$ and $a_*$ between $-0.5$ and $-0.94$.
For \sgra, we infer models that go beyond MAD/SANE with $a_* \sim 0.8$ -- 0.9, $i_\mathrm{los} \sim \ang{20}$  -- \ang{40} (corrected for GRMHD symmetry for  clockwise rotating accretion flows), $R_\mathrm{high} \sim 14$, and $\theta_\mathrm{PA} \sim \ang{106}$ -- \ang{137}.
Our results are in broad agreement with the literature, but the inferred position angle of \sgra's spin axis implies 1.3\,mm ALMA measurements are picking up significant emission from non-horizon scales.
As a next step, we plan to develop interpretable AI methods, building on the reverse engineering tests conducted in this initial study.
The goal is to unravel which predictions are driven by which visibility data points.

Finally, we showed how the planned AMT will likely lead to considerable improvements in parameter estimation accuracy for tests of models that go beyond GR.
For quantities of direct scientific interest, we can thus predict qualitative improvements from future EHT array upgrades.

\section{Data and code availability}
\label{sec:reproducibility}

The observational \ac{eht} data can be obtained from the \href{https://doi.org/10.25739/kat4-na03}{10.25739/kat4-na03} digital object identifier.
A single \texttt{data\_production.sh} script in a dedicated repository\footnote{\url{https://bitbucket.org/M_Janssen/casaeht}.} can be used to get a fully automated reduction of the observational data, producing the calibrated \ac{eht} data used in this work.  
This calibration is done with the containerized version 7.2.2 of the \rpicard{} pipeline\footnote{\url{https://bitbucket.org/M_Janssen/picard}.} (tagged as \href{https://hub.docker.com/layers/mjanssen2308/casavlbi_ehtproduction/646d6a189c01b04cfa10077a46650038d61687d9_377f3692631b8037f1ffc2bb3d1f9620bc209ca2/images/sha256-f8d8db50851fc37dc6c8b1c8a326120453147b085ed97b77bd68795374ce0b9c?context=explore}{\texttt{646d6a189c01b04cfa10077a46650038d61687d9\_\newline25c42c3c75a8334d1be4f72bc56b4344dc1f068e}}\footnote{\url{https://hub.docker.com/r/mjanssen2308/casavlbi_ehtproduction}.}).

The synthetic training data are generated with the \symba{} pipeline\footnote{\url{https://bitbucket.org/M_Janssen/symba}.} \citep{2020Roelofs}. The containerized\footnote{\url{https://hub.docker.com/r/mjanssen2308/symba}.} \texttt{\href{https://hub.docker.com/layers/mjanssen2308/symba/49a813d2dc62eac809f3909bee0d38a8b113ffc4/images/sha256-4a31d46480ffa1b6abfb55fd52fd829274aece3b4f12c5e563b31a810eb9c373?context=explore}{49a813d2dc62eac809f3909bee0d38a8b113ffc4}} \symba{} version used in this study is based on \rpicard{} version 7.2.2 and the \texttt{focalpy38.fd529fd} version of the \meqsilhouette{}\footnote{\url{https://github.com/rdeane/MeqSilhouette}.} software \citep{Blecher2017, Natarajan2021}.

The availability of the synthetic GRMHD training datasets is described in \citet{zingularity1}.
The availability of the \z{} code and configuration files needed to instantiate, train, and apply the BANNs is given in \citet{zingularity2}.

\begin{acknowledgements}

We thank Jesse Vos for useful discussions about the interpretation of our results. 

This publication is part of the M2FINDERS project which has received funding from the European Research Council (ERC) under the European Union's Horizon 2020 Research and Innovation Programme (grant agreement No 101018682).

JD is supported by NASA through the NASA Hubble Fellowship grant HST-HF2-51552.001A, awarded by the Space Telescope Science Institute, which is operated by the Association of Universities for Research in Astronomy, Incorporated, under NASA contract NAS5-26555.

MW is supported by a Ram\'on y Cajal grant RYC2023-042988-I from the Spanish Ministry of Science and Innovation.

This material is based upon work supported by the National Science Foundation under Award Numbers DBI-0735191,  DBI-1265383, and DBI-1743442. URL: \url{www.cyverse.org}.
This research was done using resources provided by the Open Science Grid, which is supported by the National Science Foundation award \#2030508.
This research used the Pegasus Workflow Management Software funded by the National Science Foundation under grant \#1664162.
Computations were performed on the HPC system Cobra at the Max Planck Computing and Data Facility
This research made use of the high-performance computing Raven-GPU cluster of the Max Planck Computing and Data Facility.
Corner plots of posteriors were created with \texttt{corner.py} \citep{corner}.

\end{acknowledgements}

\bibliographystyle{aa} 
\bibliography{link_to_init}

\end{document}